\documentclass[sigconf]{acmart}

\usepackage{graphicx}
\usepackage{makecell}
\usepackage{tabularx}
\usepackage{textcomp}
\usepackage{multirow}
\usepackage{caption}
\usepackage{subcaption}
\usepackage{array}

\AtBeginDocument{%
  }

\copyrightyear{2026}
\acmYear{2026}
\setcopyright{cc}
\setcctype{by-nc-nd}
\acmConference[W4A '26]{Web4All 2026 23rd International Web for All Conference}{June 29 -- July 03, 2026}{Dubai, United Arab Emirates}
\acmBooktitle{Web4All 2026 23rd International Web for All Conference (W4A '26), June 29 -- July 03, 2026, Dubai, United Arab Emirates}
\acmDOI{10.1145/3800424.3800450}
\acmISBN{979-8-4007-2372-8/2026/04}

\begin{document}

\title[AR for Communication Access in DHH Experiential Higher Education]{Evaluating the Feasibility of Augmented Reality to Support Communication Access for Deaf Students in Experiential Higher Education Contexts}

\author{Roshan Mathew}
\orcid{0000-0001-6618-2802}
\email{rm1299@rit.edu}
\affiliation{%
  \institution{Accessible and Immersive Realities (AIR) Lab, \\Rochester Institute of Technology}
  \streetaddress{One Lomb Memorial Drive}
  \city{Rochester}
  \state{New York}
  \country{USA}
  \postcode{14623}
}

\author{Roshan L. Peiris}
\orcid{0000-0002-4191-3565}
\email{roshan.peiris@rit.edu}
\affiliation{%
  \institution{Accessible and Immersive Realities (AIR) Lab, \\Rochester Institute of Technology}
  \streetaddress{One Lomb Memorial Drive}
  \city{Rochester}
  \state{New York}
  \country{USA}
  \postcode{14623}
}

\renewcommand{\shortauthors}{Mathew and Peiris}

\begin{abstract}
Deaf and hard of hearing (DHH) students often experience communication barriers in higher education, which are particularly acute in experiential learning environments such as laboratories. Traditional accessibility services, such as interpreting and captioning, often require DHH students to divide their attention between critical tasks, potential safety hazards, instructional materials, and access providers, creating trade-offs between safety and equitable communication. These demands can disrupt task engagement and increase cognitive load in settings that require sustained visual focus, highlighting the limitations of current approaches. To address these challenges, this study investigates Augmented Reality Real-Time Access for Education (ARRAE), an ecosystem based on augmented reality (AR) smart glasses, as a potential intervention for laboratory-based environments. By overlaying interpreters or captions directly into a student’s field of view, AR enables the integration of accessibility into hands-on learning without compromising safety or comprehension. Through an empirical study with 12 DHH participants, we evaluate how AR-mediated access influences visual attention patterns and perceived cognitive load during hands-on tasks. The findings suggest that AR-mediated communication shows strong potential to improve attention management and communication accessibility in experiential learning environments, though participants emphasized that accessibility preferences are highly context-dependent. Participants also identified several design and ergonomic challenges, including display positioning, visual fatigue, and compatibility with hearing devices. Together, these results highlight both the promise of AR for supporting accessible participation in visually demanding environments and key design considerations for future systems.
\end{abstract}



\maketitle

\section{Introduction}
Deaf and hard of hearing (DHH, henceforth deaf) students continue to experience persistent inequities in higher education \cite{batista_deaf_2023,lang_higherEd_2002} illustrated by a 15.7\% gap in degree attainment compared to hearing peers \cite{deafCenterPostsecData}. Although enrollment among deaf students has risen since the 1980s due to legislation and more accessible educational environments \cite{newman_school_outcome}, these gains have not closed the attainment gap, suggesting that structural barriers beyond students’ control hinder their educational progress. One significant barrier is the lack of accessibility within academic settings. Legal frameworks, such as the Americans with Disabilities Act (ADA) of 1990 \cite{ada_1990} and Section 504 of the Rehabilitation Act of 1973 \cite{section504} mandate equal access and reasonable accommodations in educational settings, often provided through services such as sign language interpreters and real-time captioners that render auditory information visually \cite{kushalnagar_resna_2013}. However, research shows that the presence of these services alone does not ensure equal access. A frequently cited challenge is the heavy cognitive load of processing all information visually \cite{chandler_cognitive_1991, mather_issue_2012, kushalnagar_collaborative_2014}.

In typical classroom settings, hearing students benefit from a multi-channel learning environment in which auditory information from the lecturer complements visual information from slides, demonstrations, and other classroom activities \cite{mather_issue_2012}. In contrast, the learning experience of deaf students is almost entirely visual. They must actively manage and synthesize multiple, spatially dispersed streams of visual information in real time \cite{mather_issue_2012}. This challenge is often described as the split-attention effect, which occurs when learners must integrate information presented across separate but interdependent sources, resulting in less effective learning compared to when the same information is presented in an integrated format \cite{mather_issue_2012}. For deaf students, this means continuously shifting attention between primary instructional content (e.g., an instructor’s demonstration or slide deck) and the visual translation of spoken language provided by access services \cite{kushalnagar_collaborative_2014}. Importantly, this division of attention is not a passive process; it requires sustained cognitive effort \cite{kushalnagar_collaborative_2014}. As a result, deaf students may experience information loss, reduced engagement with course materials, and a heightened cognitive burden. These factors can leave deaf students feeling less confident in their understanding and, despite accommodations, at risk of falling behind their hearing peers \cite{cavender_classinfocus_2009}. In practice-intensive environments such as laboratories, the stakes are even higher. For example, students working at biosafety cabinets or fume hoods must maintain visual focus on equipment for safety, which often means losing access to interpreters or captions \cite{smith_chemical_2016, gehret_experiential_2023, craig_navigating_2022}. The result is a trade-off between safe participation and equitable communication, an accessibility failure rooted not only in resources but in interaction design.

In response to these challenges, researchers have examined pedagogical, environmental, and technological interventions. Pedagogical strategies, such as strategic pausing or providing materials ahead of time, aim to reduce cognitive demands during real-time tasks \cite{macdonald_deaf_2002, tips_teaching_dhh, making_science_labs_accessible}. Similarly, environmental strategies optimize classroom layouts to maintain clear lines of sight \cite{mather_issue_2012}. While helpful, these approaches face significant limitations in experiential settings: chemical reactions cannot be paused, and safety protocols often strictly dictate where a student must look, making line-of-sight adjustments insufficient. Technological interventions have aimed to lessen the dispersion of visual information by consolidating multiple streams onto a single digital display or by using attention-guiding interfaces that highlight the most relevant source in real time \cite{kushalnagar_collaborative_2014, cavender_classinfocus_2009}. Augmented Reality (AR) has been explored as a promising approach to enhance content presentation while addressing line-of-sight issues and the split-attention effect \cite{mathew_aod_2022, miller_use_2017, yang_holographic_2022, yang_holographic_user_study_2022, ioannou_augmented_2018}. Collectively, these studies show that AR can improve understanding and reduce the cognitive load often associated with traditional accessibility methods. However, most AR accessibility research in education has focused on general classrooms, leaving a critical gap in addressing the safety and attention needs of laboratory environments.

Experiential learning, particularly in laboratory settings, enables students to bridge theory with practice while developing transferable skills \cite{easton_effect_2025}. Studies show its effectiveness is amplified by visual aids that align with the visual learning strengths of many deaf students \cite{anwar_effectiveness_2024}. In this work, we explore augmented reality (AR) smart glasses as a technological intervention to reduce split attention and enhance communication access in higher education, with a focus on experiential learning contexts. We introduce Augmented Reality Real-Time Access for Education (ARRAE), a platform that leverages AR smart glasses to overlay interpreters or captions directly within a deaf student’s field of view (FOV). AR has the potential to integrate accessibility into hands-on learning without compromising safety or comprehension. To assess this potential, we conducted a usability study in a simulated laboratory environment involving 12 deaf students. Participants were divided into two groups of six based on their primary communication preference in academic settings: interpreting or captioning. Each group was supported by either an American Sign Language (ASL) interpreter or a real-time captioner who translated the instructor's spoken English instructions. The following research questions guided our study:
    \begin{enumerate}
        \item How does AR-mediated communication access (interpreting and captioning) impact the visual attention patterns of deaf students during hands-on laboratory tasks, and how do these patterns correlate with their perceived cognitive load compared to traditional access methods?
        \item What usability and design feasibility requirements emerge when deploying AR smart glasses as a communication access tool in experiential higher education settings?
    \end{enumerate}
Our contributions are threefold. First, we present an empirical study of deaf students’ experiences with both traditional and AR-supported accessibility practices, with particular attention to interactional challenges such as visual dispersion. Second, we evaluate the usability and design feasibility of AR smart glasses as a platform for real-time communication access in experiential higher educational contexts. Third, we discuss design implications for developing inclusive, technology-mediated accessibility platforms that move beyond reactive accommodation toward systemic integration in educational interaction design. Overall, by centering our investigation in laboratory settings, we address the unique ergonomic and cognitive demands of high-stakes, safety-critical environments that have been largely overlooked in prior accessibility research.

\section{Related Works}

\subsection{Accessibility for Deaf Students in the Classroom and Laboratory}

Attending mainstream classes and laboratory sessions presents unique challenges for deaf students, as their learning environments are marked by extraordinary complexity and high cognitive demand. In a typical lecture, they must simultaneously process and integrate information from multiple, disparate visual channels: the instructor’s physical presence, non-verbal cues, slides or whiteboards, and linguistic information conveyed through interpreters or real-time captions \cite{olszak_learning_2025, braun_welcoming_2018}. This constant need to split visual attention is a primary academic barrier and a significant source of extraneous cognitive load \cite{braun_welcoming_2018}. 

Eye-tracking studies reveal that deaf students continually shift their gaze between interpreters, captions, slides, and instructors, resulting in information loss during the saccadic intervals \cite{pelz2008visual, agarwal2021}. This challenge intensifies in laboratories, where students must simultaneously manipulate equipment and monitor visual communication channels. The physical environment of a laboratory is often inherently hostile to auditory-based communication; loud machinery, such as fume hoods, generates background noise that renders assistive listening devices ineffective \cite{listman_inclusive_2024}. Visually, crowded lab benches can obstruct lines of sight to the instructor or interpreter. Furthermore, the confined space of many traditional labs creates ergonomic friction, making it difficult to accommodate the necessary triad of student, instructor, and interpreter \cite{listman_inclusive_2024}. Without careful coordination, critical safety instructions may be lost.

Thus, accessibility requires not only providing accommodations but also restructuring the interaction design of the learning environment to align with the reality that students cannot visually attend to multiple sources at once.

\subsection{Sign Language Interpreting Services}

Sign language interpreting remains a cornerstone of access for many deaf students. Controlled studies demonstrate that mediated instruction via highly skilled interpreters can be as effective as direct instruction from a signing instructor \cite{marschark_learning_2008}. However, the day-to-day reality in higher education is often less ideal. Even with a skilled interpreter, the student faces the cognitive load of processing information through a visual intermediary while attending to other visual stimuli \cite{braun_welcoming_2018}.

The proliferation of Video Remote Interpreting (VRI), often delivered via web-based streaming protocols, represents a significant shift in service delivery. While VRI mitigates provider shortages \cite{deaf_center_video_remote_interpreting}, it introduces technical and interactional points of failure. Reliance on standard web-conferencing setups (2D video) creates communicative hurdles: remote interpreters struggle to perceive classroom dynamics or manage turn-taking, while signers must constrain their movements to fit the camera frame \cite{alley_exploring_remote_interpreting_2012}. These factors degrade the quality of access compared to in-person interpreting, highlighting a need for display technologies that reintegrate the remote interpreter into the physical context.

\subsection{Real-time Captioning Services}

Captioning and speech-to-text services provide a vital bridge to auditory information for deaf students.  The quality divide between human and machine-generated solutions is stark. Communication Access Real-time Translation (CART) meets the ADA’s effective communication benchmark (>99\% accuracy) \cite{deaf_center_effective_comm}. In contrast, Automatic Speech Recognition (ASR) systems frequently fail in noisy classrooms, producing fragmented transcripts that force students to engage in the cognitively taxing process of error correction \cite{deaf_center_asr}.

Research on deaf students' experiences with these web-delivered text streams highlights the centrality of accuracy and user control \cite{kawas_improving_2016}. A consistent theme is the desire for independence; students seek systems they can operate and customize without relying on instructors to manage the hardware. These findings suggest that future interfaces must offer customizable displays, high accuracy, and minimal setup friction.

\subsection{Augmented Reality Smart Glasses}

AR smart glasses enter the landscape not merely as a display hardware, but as a platform to fundamentally restructure the ergonomics of accessibility. Unlike laptops or tablets that introduce an additional screen, AR integrates the accommodation directly into the user's natural FOV \cite{miller_use_2017}. By projecting web-streamed captions or interpreter video onto a semi-transparent display, the technology allows simultaneous viewing of the physical task and the digital accommodation \cite{mathew_augmented_2023}. 

This approach is grounded in Cognitive Load Theory (CLT) and the Spatial Contiguity Principle of Multimedia Learning \cite{mayer_spatial_contiguity_principle_2009}. CLT suggests that the mental integration of spatially separated information sources imposes extraneous cognitive load \cite{ayres_cognitive_load_theory_2009}. AR smart glasses mitigate this by physically overlaying the accommodation, thereby reducing the split-attention effect and freeing working memory for germane learning tasks.

A prior study investigated the use of AR glasses for deaf students in an academic context to test the hypothesis that consolidating visual information using smart glasses would improve lecture comprehension \cite{miller_use_2017}. The qualitative results from this study showed that participants overwhelmingly provided positive feedback on the experience of using the glasses. The most common theme was that the glasses significantly reduced the need to constantly shift their attention between the interpreter and the lecture slides. Users consistently reported that it was easier to follow the lecture with the smart glasses and that they could clearly see the potential for this technology to be used in a real classroom setting. 
\section{ARRAE Platform Implementation}
We developed ARRAE, a platform centered on optical see-through (OST) smart glasses, to enable deaf students to access real-time captioning and sign language interpreting. Unlike traditional access models that require shifting attention to a secondary screen, ARRAE positions real-time captions or sign language video directly within the user’s FOV. This heads-up display (HUD) approach supports the spatial contiguity principle, ensuring students remain visually engaged with hands-on laboratory activities while receiving communication access.  

ARRAE comprises three core components: an integrated web portal, a low-latency server, and an AR smart-glasses application. The web portal implements a multi-account system that supports different stakeholder roles, including deaf students, instructors, interpreters, and captioners. Through their accounts, stakeholders can manage information, schedule events, and access tools relevant to their role. The portal is accessible on any Internet-connected device, including smartphones and laptops. The server acts as a bridge between the portal and the AR application and hosts databases for user accounts and event data. The AR application, developed for Vuzix Blade 2 smart glasses \cite{vuzix_blade}, requires Wi-Fi to connect to the server and stream either real-time interpreting or captioning. We chose the Vuzix Blade 2 because it is a commercially available, lightweight, self-contained pair of AR smart glasses featuring advanced waveguide optics for hands-free connectivity. Vuzix provides a robust Android-based software development kit (SDK), which we used to develop our prototype application. For captioning, the application provides several customization features: display brightness (three levels), font size (three options), font style (four options), and font color (six options). Real-time captioning is powered by integrated TypeWell technology \cite{typewell_transcription}, which captioners can access through the portal. TypeWell was selected for this study because it is widely regarded as a gold-standard tool for real-time educational captioning.

\subsection{System Architecture}
The ARRAE ecosystem is built on a modular web architecture designed for low-latency performance and high scalability. The system utilizes Nginx as a reverse proxy to manage incoming traffic from supported devices and web browsers, routing requests to a core backend built on the Fastify Node.js framework.

\subsubsection{Web Portal and Stakeholder Interfaces}
The user interface is a Single Page Application (SPA) developed using ReactJS. This portal serves distinct views for three primary stakeholders: the Interpreting Portal (video streaming), the Captioning Portal (text streaming), and the Student View (AR output). The frontend uses a Model-View-Controller (MVC) pattern to dynamically retrieve and modify session data via a REST API.

\subsubsection{Real-Time Communication Protocols}
To ensure the synchronicity required for accessible communication, the platform employs a dual-protocol strategy:
\begin{itemize}
    \item Bi-Directional State Management: Once a session is initialized, the system transitions from standard HTTP requests to Socket.IO. This enables persistent, bi-directional communication between the server and the clients (smart glasses and web portals) to handle session state changes, such as connection handshakes and user status updates, with minimal overhead.
    \item Media Streaming (WebRTC): For high-bandwidth data, specifically the sign language interpreter’s video feed and the return feed of the lab environment, the system utilizes a WebRTC adaptor. This connects to the media server (via Ant Media's API) to deliver sub-second-latency video, which is critical for maintaining lip sync and fluid signing.
\end{itemize}

\subsubsection{Captioning Integration (TypeWell)}
Text-based access is handled through a specialized pipeline. The Fastify backend spawns a dedicated child process to function as a TypeWell ingestion server. This allows the platform to ingest captioning streams and broadcast them via the internal socket layer directly to the AR interface. This architecture isolates the captioning process, ensuring that heavy video traffic does not degrade the performance of text delivery.

\subsubsection{Event Execution}
At the start of an event, deaf students log in to retrieve event details and a QR code. Scanning the QR code with the smart glasses links the device to the assigned service stream.
\begin{itemize}
    \item Interpreting.
    Assigned interpreters log in to the portal to access event information and initiate streaming. Once streaming begins, the interpreter feed appears in the student’s display. ARRAE supports team interpreting, a standard practice where interpreters switch every 15–20 minutes \cite{rid_interpreter_resources_2022, hoza_team_2022}. 
    
    \item Captioning.
    Captioners access a dedicated captioning interface within the portal to begin transcription. They can view video streams from the laboratory to enhance contextual information and aid their captioning process. Once captioning starts, the live text stream is displayed on the student’s smart glasses.
    
    \item Monitoring.
    Instructors and providers can monitor all video feeds (student, lab, and interpreter/captioner) to ensure safety and situational awareness, while the student views only their selected access service.
\end{itemize}

\section{Method}
The study was approved by the ethics review board of the authors’ institution. We conducted a within-subjects usability study (N = 12) to assess the feasibility of ARRAE in experiential learning settings. Participants were stratified into two groups based on their primary communication preference: ASL interpreting (n = 6) and real-time captioning (n = 6). Each participant completed tasks under three conditions: (1) In-Person Access, (2) Remote Access (Traditional VRI/Captioning), and (3) AR-Mediated Access (ARRAE).

\subsection{Study Design and Environment}
The study utilized two distinct physical locations to simulate the separation often present in remote service delivery.

\subsubsection{Location A: Simulated Laboratory}
\begin{figure*}
    \centering
    \clearpage
    \setlength{\belowcaptionskip}{-20pt}
    \includegraphics[width=0.8\textwidth]{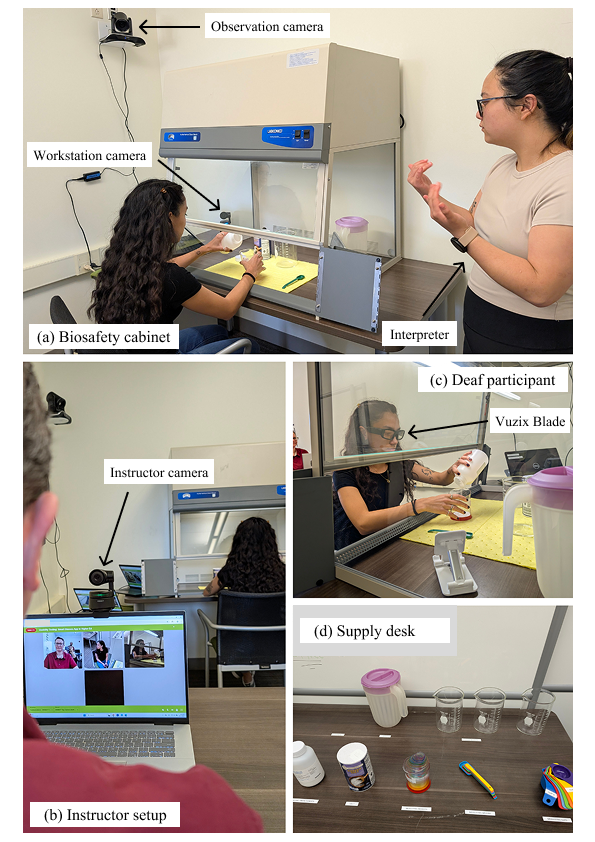}
    \Description{A representative figure showing the experimental setup used to evaluate ARRAE in a simulated laboratory environment. Subfigure (a) shows a deaf participant seated at a biosafety cabinet workstation while an interpreter stands nearby; both the workstation and observation cameras are visible. Subfigure (b) shows the instructor’s view, including a laptop displaying multiple video feeds and an instructor camera aimed at the workstation. Subfigure (c) shows a close-up of the deaf participant wearing Vuzix Blade smart glasses while performing a task. Subfigure (d) shows the supply desk with labeled materials used during the task. Together, these illustrate the physical layout, participant roles, and equipment used to support interpreting and captioning using the ARRAE platform.}
    \caption{ARRAE study setup. (a) Biosafety cabinet with workstation, observation cameras, and an interpreter present. (b) Instructor setup with laptop display and instructor camera. (c) Deaf participant wearing Vuzix Blade smart glasses while performing the task. (d) Supply desk with labeled materials. Figures do not include the real participants of the study.}
    \label{fig:location-a}
\end{figure*}

Location A simulated a laboratory where deaf participants performed the experimental tasks (see Figure \ref{fig:location-a}). A biosafety cabinet/fume hood was positioned facing a wall, with the instructor’s desk located several meters behind it, intentionally placed out of the participant’s direct line of sight to simulate typical laboratory communication barriers. A separate desk, located near the instructor’s station, held study materials that participants collected during tasks, allowing us to evaluate the access modalities when mobility in the lab was required. 

To enable remote observation, the workstation was instrumented with two cameras: an OBSBOT Tail Air 4K PTZ camera \cite{obsbot_tail_air} streaming the participant’s hands/workstation, and an overhead Tenveo 10x Zoom PTZ camera \cite{tenveo_vhd10u} capturing the participant and their signing space. Both feeds were streamed to the ARRAE web portal, providing the instructor and access providers with contextual views of the interpreting or captioning environment. The instructor’s station included an OBSBOT Tiny 4K webcam \cite{obsbot_tiny} to stream the instructor and their spoken instructions. The OBSBOT Tail Air, OBSBOT Tiny, and the Tenveo 10x Zoom PTZ cameras were chosen based on the findings from a prior study where interpreters reported a preference for PTZ cameras in remote interpreting contexts \cite{mathew_improving_2025}. 

In the In-Person Access condition, access service providers were positioned beside the participant. In the Remote Access condition, participants used an iPad in a transparent protective sleeve inside the fume hood to access standard VRI or captioning services. In the ARRAE condition, participants viewed the interpreter or captions using Vuzix Blade 2 smart glasses. The iPad was also used to complete study surveys. This setup enabled comparison of AR-mediated access with traditional service methods in laboratory environments.

\subsubsection{Location B: Remote Service Node}
\begin{figure*}
    \centering
    \clearpage
    \setlength{\belowcaptionskip}{-10pt}
    \includegraphics[width=0.8\textwidth]{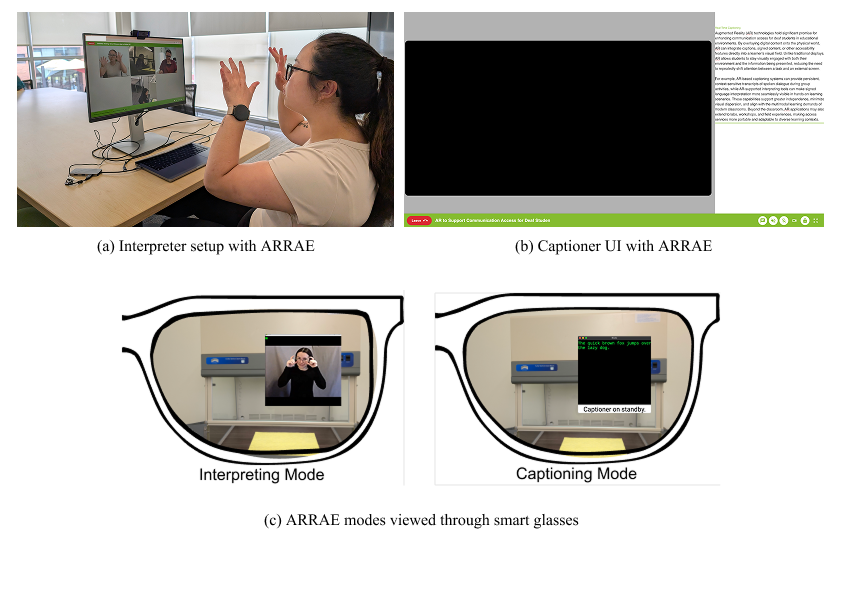}
    \Description{This figure shows the interfaces and display modes used by interpreters, captioners, and deaf participants while using the ARRAE platform. Subfigure (a) depicts the interpreter’s setup at Location B, with multiple video feeds displayed on a monitor while the interpreter signs toward the camera. Subfigure (b) shows the captioner’s user interface in ARRAE, where typed captions are entered and transmitted to participants. Subfigure (c) illustrates the two modes as they appear through AR smart glasses: in interpreting mode, the interpreter’s video feed is overlaid in the participant’s field of view; in captioning mode, real-time text captions are displayed instead. Together, these figures demonstrate how ARRAE supports interpreting and captioning through dedicated provider interfaces and participant-facing AR displays.}
    \caption{ARRAE interfaces and display modes. (a) Interpreter setup at Location B showing video feeds of the participant's workspace. (b) Captioner user interface with ARRAE. (c) Smart glasses view illustrating the interpreting mode and captioning mode. Figures do not include the real participants of the study.}
    \label{fig:location-b}
\end{figure*}

Location B served as the remote hub (see Figure \ref{fig:location-b}). It was equipped with a laptop and an external monitor displaying the multi-angle feeds from Location A. During Remote Access and ARRAE conditions, the interpreter and captioner worked from this location, utilizing the ARRAE web portal to view the student and the lab environment while streaming their services back to the student's iPad or smart glasses.

\subsubsection{Support Personnel}
To ensure that all deaf participants in the interpreting and captioning groups experienced their tasks under identical conditions, only one instructor, one interpreter, and one captioner were employed throughout the study. The instructor had 10–15 years of experience teaching STEM courses in higher education, using spoken English as the primary language of instruction. The interpreter was a certified ASL interpreter with 3–5 years of professional experience in higher education settings. This included in-person interpreting as well as less than one year of experience providing VRI in higher education contexts. The captioner had completed a comprehensive employer-sponsored training program to acquire the necessary skills and knowledge for effective captioning. They had 5–10 years of professional captioning experience, including 5–10 years of in-person real-time captioning in higher education settings. They also had 3–5 years of experience providing remote real-time captioning in higher education. We note that relying on a single individual for each role may have introduced the limitation that their specific performance may have influenced the outcomes; however, this approach was considered appropriate to balance ecological validity with experimental control.

\subsection{Study Participants}
To be eligible, participants were required to identify as D/deaf, hard of hearing, DeafBlind, oral-deaf, or late-deafened \cite{nad_definitions}, and to have prior experience using either interpreting or real-time captioning as access services in higher education settings. They were also required to be 18 years of age or older. We employed purposeful and snowball sampling via email, social media, and community outreach to recruit eligible individuals.

\subsection{Participant Demographics}
Six participants were recruited for the interpreting study group. Ages were reported in ranges: one participant was 18–20 years old, two were 21–24 years old, two were 25–29 years old, and one was 35–39 years old. Four participants identified as female and two as male. In terms of ethnicity, four identified as White or Caucasian, one as Asian, and one as multiracial. In terms of education, one participant was a high school graduate pursuing a bachelor’s degree, four had completed a bachelor’s degree, and one held a master’s degree. With respect to hearing status, four participants identified as Deaf and two as hard of hearing. Four reported not using hearing devices, while two used bilateral hearing aids (HAs). All participants considered themselves native ASL users, preferred to receive information from hearing individuals through ASL, and indicated that ASL interpreting was their most preferred access services method in classrooms. Participants’ receptive ASL proficiency was assessed using American Sign Language Comprehension Test (ASL-CT) \cite{hauser_american_2016}. Scores ranged from 17 to 28 out of 30 (M = 23.5, SD = 3.78, Mdn = 24.0). These results indicate that most participants demonstrated intermediate to advanced comprehension, with one participant scoring lower but still meeting the inclusion criteria.

Six participants were also recruited for the captioning study group. Ages were collected in ranges: three participants were 21 - 24 years old, two were 30 - 34 years old, and one was 35 - 39 years old. Three participants identified as female, two as male, and one as non-binary. Regarding ethnicity, four participants self-identified as White or Caucasian, one as Asian, and one as multiracial. In terms of education, one participant had an Associate's degree, three participants had a Bachelor's degree, and two participants had a Master's degree. In terms of hearing status, two participants self-identified as Deaf, two as deaf, and two as hard of hearing. Hearing device use varied: one participant did not wear any hearing devices, two used bilateral HAs, one used a Cochlear Implant (CI) in one ear, one used bilateral CIs, and one used an HA in one ear and a CI in the other ear. In terms of communication preferences, all six participants indicated that reading captions was their most preferred way of accessibility accommodation in classrooms. For both these study groups, we have omitted specific demographic information to protect participant anonymity. Silent reading fluency was assessed with the Test of Silent Contextual Reading Fluency - 2 (TOSCRF-2) \cite{toscrf2_proed}. Raw scores ranged from 99 to 174 (M = 140.5, SD = 28.9, Mdn = 140.0), indicating variability across participants but generally high levels of silent reading proficiency for an adult sample.

\subsection{Study Procedure}
Sessions were moderated by deaf researchers fluent in the participants' preferred communication method (ASL interpreting or captioning), ensuring cultural competence and minimizing linguistic barriers. This enhanced data accuracy, encouraged candid feedback, and provided contextually grounded insights into participants' experiences with different access modalities. Each 90-minute session proceeded as follows:

\textbf{1. Training (30 min):} Participants were briefed on the study’s purpose, completed informed consent, and received training with the Vuzix Blade 2 smart glasses to familiarize themselves with the ARRAE system and navigation. Once participants confirmed their understanding, they proceeded to the second phase. 

\textbf{2. Experimental Tasks (30 min):} Participants completed three hands-on tasks guided by spoken English instructions. The task required making artificial snow using an instant snow polymer; the measurements and materials varied across tasks to mitigate learning effects and ensure reliance on accurate instructions. Each task was paired with a different access modality (in-person, remote, or ARRAE), and to control for order effects, the three access conditions were counterbalanced across tasks. Post-task questionnaires were administered immediately after each condition.

\textbf{3. Evaluation (30 min):} Participants completed a comparative survey reflecting on their experiences with the three access modalities, followed by a demographic survey. Participants were then given a five-minute break, followed by a semi-structured interview facilitated by a moderator. Interviews focused on usability, feasibility, and satisfaction with receiving access services through smart glasses, as well as comparative benefits and barriers relative to in-person and remote modalities. In the captioning study, the captioner remained to provide captioning support during interviews. Finally, participants completed either the ASL--CT to assess receptive ASL skills or the TOSCRF--2 to assess English reading fluency. These standardized measures were implemented to determine whether participants' ASL competency or reading comprehension affects the outcome measures. Each session concluded with a debrief.

\subsection{Data Analysis}
We used descriptive statistics to summarize participant demographics. Prior to analysis, the distribution of rating variables was assessed using the Shapiro-Wilk test for normality. Because the rating data were ordinal and the sample size was small (n = 6), Friedman tests were used to examine differences in participant ratings across study conditions in SPSS Statistics \cite{spss_ibm_2024}. When omnibus tests were significant, we conducted post hoc pairwise comparisons with Bonferroni correction to control for inflated Type I error.

As the interviews with deaf participants in the interpreting study were conducted in ASL, the deaf researcher reviewed the recordings and generated ASL-English transcripts, which were subsequently reviewed for accuracy by certified ASL interpreters affiliated with the research team who were not present during the study sessions. In the captioning study, transcripts were produced in real time by the captioner and later reviewed against the recordings for accuracy. We analyzed interview and open-ended responses using Braun and Clarke’s reflexive thematic analysis approach \cite{braun_using_2006}. To ensure reliability and mitigate individual bias, researchers independently reviewed the transcripts and generated initial codes. They then compared codes, discussed discrepancies, and refined the thematic framework until consensus was reached. Throughout this process, researchers maintained reflexive memos to document positionality and analytic decisions. Preliminary themes were further shared with the broader research team for peer debriefing, providing additional opportunities to question assumptions. To support our analysis, we include excerpts from participants that illustrate key themes and demonstrate the alignment between the data and our interpretations.

\section{Findings}

\subsection{Interpreting Study}

\subsubsection{Quantitative Results}

\paragraph{Line of Sight}
Participants' satisfaction with the line of sight was evaluated for three interpreting modalities: in-person, VRI, and ARRAE. Ratings were captured on a 100-pt visual analog scale (VAS) anchored by \textit{Completely Unsatisfied} (1) and \textit{Completely Satisfied} (100). Median satisfaction scores were slightly higher for in-person interpreting ($Mdn = 82.00, IQR = 76$) followed by VRI ($Mdn = 79.00, IQR = 31$) and ARRAE ($Mdn = 76.50, IQR = 40$). Although in-person interpreting received the highest median rating, responses for this condition showed greater variability. A Friedman test confirmed that the observed differences across the three interpreting conditions were not statistically significant, $\chi^2(2) = 0.33, p = .846$.

\paragraph{Visual Dispersion}
Participants rated how often they had to look away from their hands-on tasks to receive information via each modality. Ratings were captured on a 100-pt VAS anchored by \textit{Very Rarely} (1) and \textit{Very Frequently} (100). While median scores showed a slight decrease from VRI ($Mdn = 75.50, IQR = 35$) to ARRAE ($Mdn = 67.50, IQR = 32$) and in-person interpreting ($Mdn = 62.00, IQR = 48$), the differences were not statistically significant, $\chi^2(2) = 0.33, p = .846$.

\paragraph{Comparing Gaze for Interpretation Delivery}
To assess the perceived benefit of receiving information in their line of sight through the AR smart glasses, participants rated their agreement with the statement: \textit{"the smart glasses allow me to receive critical information wherever I gaze primarily at."} This assessment was conducted in three contexts: a comparison of smart glasses to in-person interpreting, a comparison to a mobile display (VRI), and an evaluation of the smart glasses on their own. All ratings were captured on a 100-pt VAS. A Friedman test revealed a statistically significant difference in how participants rated the benefits of gaze via the smart glasses across the three scenarios, $\chi^2(2) = 9.24, p = 0.010$. Post-hoc analysis with a Bonferroni correction revealed that ratings for using smart glasses on their own ($Mdn = 87.00$) were significantly higher than when compared to VRI ($Mdn = 41.00$) ($p = .018$). No significant differences were found between comparisons to in-person interpreting ($Mdn = 83.50, p = 1.000$) nor between VRI and in-person interpreting ($p = .130$).

When evaluated on their own merits, the smart glasses were perceived positively, with five of six participants rating their agreement above 70. However, comparisons to VRI showed the most variability, indicating inconsistent views on the benefits relative to tablet-based remote interpretation. Participant IP05's ratings were particularly notable, rating the smart glasses at 0 for comparative contexts but highly as a standalone tool.

\paragraph{Future Use}
Participants rated their likelihood of using each of the three interpreting modalities in future settings (1 = \textit{Very Unlikely}, 100 = \textit{Very Likely}). Although median scores indicated the highest likelihood of use for in-person interpreting ($Mdn = 69.00, IQR = 76$), followed by ARRAE ($Mdn = 53.50, IQR = 68$) and VRI ($Mdn = 37.50, IQR = 30$), a Friedman test confirmed these differences were not statistically significant, $\chi^2(2) = 0.33, p = .846$.

\subsubsection{Qualitative Themes}
The following themes outline the major findings derived from the thematic analysis of post-task interviews with deaf participants.

\paragraph{Managing the Cognitive Cost of Attention Switching}
Participants constantly weighed the cognitive cost of shifting their attention against the benefit of comprehension. In-person interpreting was consistently favored for complex tasks, as it minimized this cost; the co-located nature of the interpreter enabled seamless clarification. As IP01 explained, \textit{“I felt like there was a lot of information... so I was able to ask [the in-person interpreter] and confirm that what I was doing was correct.”} In contrast, VRI imposed significant attentional and physical costs, requiring participants to move back and forth between the workstation and the iPad. IP06 described this friction: \textit{"I had to walk all the way back to the iPad... I feel like the VRI was the slowest by far."} 

Smart glasses presented a nuanced case. For stationary activities, the heads-up display enabled parallel processing. IP06 noted, \textit{“With the smart glasses, I could remain focused the whole time... I can listen and work at the same time.”} However, during mobile tasks, the visual overlay competed for attentional resources, increasing the cognitive overhead of navigation. IP06 elaborated: \textit{"If I have the glasses on when I am walking around... it feels like I have to do 3, 4, 5 things at the same time."}

\paragraph{Need for Spatial Control and Stability}
Participants expressed a critical need for user-defined spatial control of the interpreter's video feed. A recurring request was for \textit{“picture-in-picture (PIP)”} positioning rather than a fixed center view. Furthermore, the system's default head-locked rendering created a kinesthetic conflict with Deaf conversational norms, particularly backchanneling. Instinctive head nods caused the video feed to bob, leading to dizziness. IP05 explained: \textit{“I would nod and the interpreter would move up and down... That made me dizzy. I wish I could fix the interpreter in place... It felt unnatural.”} This highlights the need for world-locked anchoring options to support embodied communication practices.

\paragraph{Visual Ergonomics}
Participants reported physical discomfort due to environmental glare and optical limitations. IP06 described a "double lens" effect when looking through the fume hood while wearing the glasses. Physiological strain also resulted from focal depth mismatch, forcing participants to refocus constantly between the 2D digital interpreter and the three-dimensional (3D) physical task. IP05 noted, \textit{“I had to work harder to focus on the interpreter... My eyes got tired, and I had a headache.”} Challenges were compounded for participants who required prescription lenses, which were not integrated into the AR smart glasses used for the study. Participants consistently contrasted this with the perceptual richness of in-person communication.

\paragraph{Interactional Transparency and Feedback}
Participants emphasized that interactional transparency, awareness of what is visible to the other party, was crucial. In-person interpreting afforded common ground for rapid micro-coordination. In contrast, mediated systems often fractured this shared space. A primary issue was the lack of visual cues, leaving participants unsure whether the remote interpreter could see them signing. IP02 described the moment of realizing the interpreter could not see them, forcing a conscious attention shift to the separate observation camera: \textit{"That was the moment of realization where I had to shift my attention... Compared to looking at in-person interpreting, [where] the process of dialogue becomes easy... I did not realize that the observation camera was set up just precisely for the interpreters to see me."} Participants expressed a desire for self-view and status indicators to restore this transparency.

\subsection{Captioning Study}

\subsubsection{Quantitative Results}

\paragraph{Line of Sight}
We assessed satisfaction with line of sight across three captioning methods using a 100-point VAS. Median scores were highest for in-person captioning ($Mdn = 81.00, IQR = 15$), followed by remote captioning ($Mdn = 79.50, IQR = 21$) and ARRAE ($Mdn = 79.50, IQR = 26$). A Friedman test confirmed these differences were not statistically significant, $\chi^2(2) = 0.64, p = .727$.

\paragraph{Visual Dispersion}
Participants rated how often they looked away from tasks to receive captions. A Friedman's test revealed a statistically significant difference across modalities, $\chi^2(2) = 9.48, p = .009$. Median scores were lowest for ARRAE ($Mdn = 55.50, IQR = 31$), followed by remote captioning ($Mdn = 75.00, IQR = 37$) and in-person captioning ($Mdn = 80.00, IQR = 30$). Post-hoc pairwise comparisons with Bonferroni correction indicated that visual dispersion was significantly lower for ARRAE than for remote captioning ($p = .042$) and in-person captioning ($p = .018$).

\paragraph{Comparing Gaze for Captioning Delivery}
Participants rated agreement with the statement, \textit{"The smart glasses allow me to receive critical information wherever I gaze,"} across three contexts. The median score was highest when comparing ARRAE to in-person captioning ($Mdn = 86.50, IQR = 24$), followed by the comparison to remote captioning ($Mdn = 84.00, IQR=31$) and the standalone evaluation ($Mdn = 84.00, IQR = 34$). A Friedman test found no statistically significant difference across these scenarios, $\chi^2(2) = .43, p = .807$.

\paragraph{Future Use}
Participants rated their likelihood of using each modality in the future. A Friedman's test revealed a statistically significant difference, $\chi^2(2) = 8.33, p = .016$. Participants rated ARRAE ($Mdn = 85.00, IQR=25$) as the most likely for future use, followed by in-person ($Mdn = 76.00, IQR=35$) and remote captioning ($Mdn = 67.50, IQR=13$). Pairwise comparisons showed participants were significantly more likely to use ARRAE than remote captioning ($p = .012$).

\subsubsection{Qualitative Themes}
This section outlines the major themes derived from the thematic analysis of post-task interviews with deaf participants in the captioning group.

\paragraph{Enhanced Mobility and Reduced Attentional Switching}
A major advantage of the smart glasses was the ability to receive information directly in the line of sight, enhancing mobility. Participants contrasted this with the inefficiency of carrying an iPad for remote or in-person captioning. CP02 explained, \textit{"I didn't have to look away as often... I didn't have to carry the iPad with me... I could double check and triple check without having to go back and forth."} The integration of information into the field of view created a sense of presence. CP04 noted, \textit{"I could see the closed captioning with the real world at the same time. So I didn't feel like I was missing anything."} Participants also valued the persistence of text in AR compared to the transient nature of Zoom captions, which disappeared too quickly for review.

\paragraph{Ergonomic and Visual Challenges}
Nearly every participant using hearing devices reported physical conflict with the glasses; the arms of the glasses often interfered with CIs or HAs. CP01 stated, \textit{"They're bulky around the ears so difficult with my hearing aids."} Visual discomfort was also reported, including blurriness and double vision caused by the difficulty of merging the digital overlay with the physical world. CP04 described this: \textit{"Trying to focus on both things at the same time sort of made the vision overlap... made it out of focus."} Participants suggested slimmer hardware and binocular displays to mitigate these issues.

\paragraph{The Role of Context in Captioning Modality Preference}
Preference was highly context-dependent. Participants distinguished between active, hands-on environments (lab) and passive environments (lecture). Smart glasses were strongly preferred for active tasks. However, in traditional lectures, the familiar iPad with in-person captioning was often seen as superior due to the ability to view large blocks of persistent text for note-taking. CP01 articulated, \textit{"In the classroom, I would stick to the iPad and smart glasses in the lab."} This underscores that the value of AR captioning is defined by the specific demands of the immediate task.
\section{Discussion}
Our study demonstrates both the promise and the current limitations of AR for accessibility in experiential learning environments. While ARRAE did not consistently outperform existing modalities across all measures, it demonstrated meaningful benefits in specific contexts, particularly when mobility and sustained task engagement were required. The study also uncovered significant interactional and ergonomic friction points that must be addressed. We interpret these findings to identify key implications for the design of AR accessibility systems.

\subsection{Spatial Contiguity as a Safety-Critical Accessibility Principle}
The primary theoretical contribution of this work is the empirical validation of the spatial contiguity principle \cite{mayer_spatial_contiguity_principle_2009} in the context of real-time accessibility. Our quantitative data, particularly for the captioning group, showed a significant reduction in visual dispersion when using smart glasses compared to remote screens. Qualitatively, participants described this as eliminating the need to shift gaze between the task and the interpreter or captions, which many found exhausting. In the laboratory context, where safety relies on continuous visual monitoring of chemical reactions or machinery, this reduction in gaze-shifting is not merely a matter of convenience; it is a matter of safety. Participants also reported a higher sense of situational awareness with AR because the transparent display allowed them to maintain peripheral vision of the room, unlike looking down at an iPad. This suggests that future accessibility standards for hazardous environments should prioritize HUDs over auxiliary screens, positioning accessibility information within the safety-critical field of view.

\subsection{The Paradox of Head-Locked Accessibility}
While the position of the display was advantageous, its behavior revealed a critical design paradox. Current AR accessibility implementations, including ARRAE, typically use a "head-locked" rendering model (where the video/text moves with the head). Our findings indicate that this model directly conflicts with the embodied nature of sign language communication. Deaf participants continuously use head nods as backchanneling signals to indicate understanding. In a head-locked system, nodding causes the interpreter’s video to bob, inducing dizziness and visual fatigue in some people. This creates a functional conflict: the very behavior used to signal \textit{"I understand"} actively degrades the ability to see the information. Therefore, users preferred to have greater spatial control over caption and interpreter displays, such as picture-in-picture positioning, to better align accessibility information with their workflow. Prior research similarly emphasizes the importance of customizable interfaces for accessibility technologies \cite{kawas_improving_2016}. This finding implies that future AR accessibility frameworks must move beyond simple HUD overlays to world-locked or body-locked stabilization, where the content remains stable in 3D space even as the user nods or turns their head.

\subsection{Interactional Transparency and Mutual Visibility in Remote Interpreting}
A recurring theme in our interviews was the loss of interactional transparency. In face-to-face settings, deaf students rely on mutual gaze to know if they are being seen. In our AR setup, students could see the interpreter, but they lacked confidence that the interpreter could see them via the Observation or Workstation cameras, echoing prior findings that remote interpreting systems can obscure shared awareness between participants \cite{alley_exploring_remote_interpreting_2012}. This asymmetry forced students to cognitively manage the camera angles, \textit{"Can the interpreter see my hands when I am signing?"}, adding an extraneous cognitive load that does not exist in in-person interactions. For AR to be a viable alternative to in-person interpreting, the system must provide explicit feedback mechanisms. Future AR interfaces could include a small self-view PIP or a status indicator (e.g., a green border) to confirm that the student's hands are currently visible to the remote provider. This would offload the cognitive burden of visibility management from the student to the interface.

\subsection{Hardware Compatibility as an Accessibility Equity Challenge}
Our study highlighted a severe hardware inclusion failure: the physical incompatibility between current AR glasses and assistive hearing devices. Nearly every participant with CIs or Behind-the-Ear (BTE) HAs reported physical conflict with the Vuzix Blade 2's arms. These challenges reflect broader limitations of current AR hardware and reinforce prior observations that accessibility applications must account for the diverse assistive technologies already used by deaf individuals \cite{miller_use_2017, mathew_augmented_2023}. This finding poses a challenge for hardware manufacturers: AR wearable designs must account for the real estate already occupied by assistive devices. Future form factors might need to decouple the display from the computing unit (e.g., lightweight clip-on optics) to accommodate the diverse range of distinct physical technologies that deaf individuals already wear.

\subsection{Task Context and Modality Preferences in Accessible Learning Environments}
Finally, our results dispel the notion that AR is a universal replacement for traditional accommodations. Participants expressed a clear preference: AR for hands-on activities in labs, and iPads/Laptops for lectures that are typically information-dense. In lectures, the stability, screen size, and persistence of text on an iPad outweigh the benefits of a HUD. In labs, the mobility afforded by AR HUDs is paramount. This suggests that university accessibility services should not view AR as a blanket solution, but rather as a specialized tool in a multimodal toolkit. The ideal accessibility ecosystem is one where a student can seamlessly transition their service stream from a laptop in the lecture hall to smart glasses in the laboratory without interrupting the session.
\section{Study Limitations and Future Directions}
We note several limitations and potential future directions in our study.

First, the sample size was intentionally small ($N=12$). While this approach generated rich qualitative data suited for an exploratory usability study, it limits the generalizability of our quantitative findings. Future work should recruit a larger, more diverse sample to validate these patterns across broader deaf populations, including those with varying levels of technological proficiency.

Second, the study utilized a single instructor, interpreter, and captioner to prioritize internal validity by keeping the delivery of instruction and access services consistent across conditions and reducing variability from differences in interpreting style or teaching pace. However, this approach may introduce bias, as the specific skills and rapport of these providers could have influenced participant experiences. Additionally, the study was conducted in a controlled laboratory with simulated safety-critical tasks, which may not fully reflect the complex acoustic and visual dynamics of a real, semester-long course. The novelty of wearing smart glasses may have also influenced satisfaction ratings. Future work should therefore include replication with multiple professionals and longitudinal, real-world deployments to assess generalizability and examine how perceptions evolve over time.

Lastly, we encountered several hardware limitations with the Vuzix Blade 2. In addition to the display resolution and head-locked rendering issues discussed earlier, the physical form factor proved incompatible with many assistive listening devices. Several participants experienced discomfort or fit issues when wearing the glasses alongside BTE HAs or CI processors, highlighting a critical accessibility gap in current AR hardware. Future research should explore next-generation AR platforms with adjustable form factors or clip-on displays that better accommodate existing assistive technologies. We plan to extend the ARRAE platform to address these ergonomic limitations and to implement the design implications identified in this study, such as shared awareness cues. We also plan to conduct a follow-up study grounded in CLT to examine whether reductions in visual dispersion translate into improved learning outcomes.

As a broader future direction, we aim to explore presenting additional aspects of speech and interpretation, including affective components of speech. For example, affective captions have been shown to enhance the user experience~\cite{fuzzyCHI, caluaCHI23, captionRoyaleCHI, hapticCaptioningCHI23, de2025tactile, cucapASSETS2025}; therefore, presenting such information through AR could further improve accessibility and enrich the overall communication experience for users.

\section{Conclusion}
This study examined the feasibility of AR to support communication access for deaf students in experiential higher education contexts. We introduced ARRAE, a platform leveraging optical see-through smart glasses, and compared it against in-person and remote interpreting and captioning in a simulated laboratory. 

Our findings suggest that while AR platforms are not yet a seamless replacement for established modalities, they offer unique affordances for hands-on environments: specifically, significant reductions in perceived visual dispersion and enhanced mobility. However, these benefits are currently tempered by ergonomic friction, visual fatigue, and a lack of support for the embodied, bidirectional nature of signed communication. Notably, our results demonstrate that accessibility is not a one-size-fits-all solution; preferences are highly context-dependent, with AR excelling in the lab and tablets preferred for lectures. For AR to become a viable component of the accessibility ecosystem, future designs must move beyond simple information overlays to prioritize physical compatibility with assistive devices, world-locked stabilization, and interactional transparency. By addressing these challenges, AR has the potential to shift accessibility from a reactive accommodation to an integrated, equitable component of the learning environment.

\begin{acks}
    The contents of this paper were partially developed under a grant from the National Institute on Disability, Independent Living, and Rehabilitation Research (NIDILRR grant number 90IFRE0083). NIDILRR is a Center within the Administration for Community Living (ACL), Department of Health and Human Services (HHS). The contents of this paper do not necessarily represent the policy of NIDILRR, ACL, or HHS, and you should not assume endorsement by the Federal Government. The authors thank Ms. Wendy Dannels for her guidance and valuable contributions to the development of this study.   
\end{acks}



\end{document}